\documentclass[psfig]{mn2e}
\usepackage{psfig}
\newcommand{\etal}{{et al.~}}

\newcommand{\Msun}{\>{\rm M_{\odot}}}

\newcommand{\beq}{\begin{equation}}
\newcommand{\eeq}{\end{equation}}

\def\Msun{\,\rm M_{\odot}}

\newdimen\hssize
\newdimen\hdsize 
\begin{document}
            
%%%%%%%%%%%%%%%%%%%%%%%%%%%%%%%%%%%%%%%%%%%%%%%%%%%%%%%%%%%%%%%%%%%%%%%%%%

\title[Profiles of dark halos]
      {On the origin of cold dark matter halo density profiles}        
\author[]
      {Yu Lu$^{1}$\thanks{E-mail: luyu@astro.umass.edu},  
       H.J. Mo$^{1}$, Neal Katz$^{1}$, Martin D. Weinberg$^{1}$
\\
       $^1$ Department of Astronomy, University of Massachusetts,
       Amherst MA 01003-9305, USA}

%%%%%%%%%%%%%%%%%%%%%%%%%%%%%%%%%%%%%%%%%%%%%%%%%%%%%%%%%%%%%%%%%%%%%%%%%%

\date{}

\pagerange{\pageref{firstpage}--\pageref{lastpage}}
\pubyear{2005}

\maketitle

\label{firstpage}

%%%%%%%%%%%%%%%%%%%%%%%%%%%%%%%%%%%%%%%%%%%%%%%%%%%%%%%%%%%%%%%%%%%%%%%%%%

\begin{abstract}
$N$-body simulations predict that CDM halo-assembly occurs in two phases: 
1) a fast accretion phase with a rapidly deepening potential well; and 
2) a slow accretion phase characterised by a gentle addition of mass to 
the outer halo with little change in the inner potential well. We 
demonstrate, using one-dimensional simulations, that this two-phase 
accretion leads to CDM halos of the NFW form and provides physical insight 
into the properties of the mass accretion history that influence the final 
profile. Assuming that the velocities of CDM particles are effectively 
isotropised by fluctuations in the gravitational potential during the fast 
accretion phase, we show that gravitational collapse in this phase leads 
to an inner profile $\rho(r)\propto r^{-1}$. Slow accretion onto an
established potential well leads to an outer profile with
$\rho(r)\propto r^{-3}$.  The concentration of a halo is determined by the 
fraction of mass that is accreted during the fast accretion phase. Using 
an ensemble of realistic mass accretion histories, we show that the model 
predictions of the dependence of halo concentration on halo formation time, 
and hence the dependence of halo concentration on halo mass, and the 
distribution of halo concentrations all match those found in cosmological 
$N$-body simulations.  Using a simple analytic model that captures much of 
the important physics we show that the inner $r^{-1}$ profile of CDM halos 
is a natural result of hierarchical mass assembly with a initial phase of
rapid accretion.  
\end{abstract}

%%%%%%%%%%%%%%%%%%%%%%%%%%%%%%%%%%%%%%%%%%%%%%%%%%%%%%%%%%%%%%%%%%%%%%%%%%


%%%%%%%%%%%%%%%%%%%%%%%%%%%%%%%%%%%%%%%%%%%%%%%%%%%%%%%%%%%%%%%%%%%%%%%%%%

\section{Introduction}
In the cold dark matter (CDM) paradigm of structure formation, most of the 
cosmic mass is locked in virialised clumps called dark matter halos. 
Luminous objects, galaxies and clusters of galaxies, are assumed to form in 
the potential wells of these dark matter halos. High resolution $N$-body 
simulations have shown that the density profiles of CDM halos can be fairly 
well described by a universal form,
\begin{equation}\label{eq:NFW}
\rho_{\rm NFW}(r)={4\rho_s \over (r/r_s)(1+r/r_s)^2}\,,
\end{equation}\label{eq:nfw}
where $r_s$ is a characteristic radius and $\rho_s$ is a characteristic 
density (Navarro, Frenk \& White 1996, 1997; hereafter (NFW). The value of 
$r_s$ is often given in units of the virial radius and one over that value 
is referred to as the halo concentration. There is still uncertainty about 
the exact value of the inner slope. While some simulations indicate that 
the inner logarithmic slope may be steeper than the NFW value, $-1$ 
(e.g. Moore \etal 1999, Ghigna \etal 2000; Fukushige \& Makino 1997, 2001,
2003), others give slopes shallower than $-1$ (Subramanian, Cen \& Ostriker
2000; Taylor \& Navarro 2001; Ricotti 2003). Jing \& Suto (2000) found that 
CDM halo profiles in their simulations do not have a completely universal
form, with the inner slope changing from system to system.

Until now, there has not been a natural explanation for the approximate 
`universal' profile resulting from the gravitational collapse of a CDM 
density field. The pioneering work by Gunn \& Gott (1972) considered the 
collapse of uniform spherical perturbations of collisionless cold dark 
matter in an expanding background. This simple model explains some 
properties of virialised objects, such as their mean density and size, but 
does not describe the density profile of a collapsed object. Subsequently, 
the spherical collapse model was extended to incorporate realistic initial 
perturbations (e.g. Gott 1975; Gunn 1977; Fillmore \& Goldreich 1984; Bertschinger 1985;
Hoffman \& Shaham 1985; Ryden \& Gunn 1987; Ryden 1988; Zaroubi \& Hoffman
1993). Bertschinger (1985) and Fillmore \& Goldreich (1984) found that halos 
have similar asymptotic inner density profiles with 
$\rho(r)\propto r^{-\gamma}$ and $\gamma\sim 2$, for a wide range of initial 
density perturbations. Unfortunately, the inner slope predicted by such 
models is much steeper than that found in 3-dimensional cosmological simulations.
A number of authors (e.g. White \& Zaritsky 1992; Ryden 1993; Sikivie \etal
1997; Subramanian 2000; Subramanian \etal 2000; Hiotelis 2002; 
Le Delliou \& Henriksen 2003; Shapiro \etal 2004; Barnes \etal 2005) noticed 
that tangential motions of particles may cause flattening of the inner 
profiles. $N$-body simulations by Huss \etal (1999) and Hansen \& Moore (2004) 
showed that  the inner density profile of a halo is correlated with the 
degree of velocity anisotropy. However, none of these has provided clear 
dynamical mechanism for the origin of the `univeral' $\rho(r)\propto r^{-1}$ 
inner profile observed in cosmological $N$-body simulations.

Lynden-Bell (1967) proposed that, as long as the initial condition for the 
collapse is clumpy, a final equilibrium state with a universal profile may 
be achieved as a result of violent relaxation that may cause a complete 
mixing of particles in phase space. Numerical simulations have shown that 
such collapses indeed produce a universal inner profile (van Albada 1982), 
but the relaxation is incomplete, in the sense that there is still a 
significant correlation between the final and initial states of a particle.
Tremaine et al. (1986) provide statistical constraints on equilibria 
resulting from violent relaxation.  In the cosmic density field predicted 
by a CDM model, the perturbations responsible for the formation of dark 
matter halos are expected to be clumpy, and so violent relaxation is expected 
to play some role in the formation of dark matter halos.  However, as shown 
by $N$-body simulations, the density profiles of virialised dark matter halos 
{\it do} depend on the formation histories of dark halos 
(e.g. NFW; Klypin \etal 2001; Bullock et al. 2001; Eke et al. 2001; Wechsler \etal 2002; Zhao \etal
2003a,b; Tasitsiomi et al. 2004; Diemand \etal 2005). NFW conjectured that
the dependence of halo concentration parameter on halo mass could be explained 
by the assembly of more massive halos at later times than lower-mass halos. 
Later, Wechsler \etal (2002) and Zhao \etal (2003a;b) found that the 
concentration of a halo depends on the assembly history. Together, these 
results suggest that CDM initial conditions play an important role in 
determining halo density profiles.

In this paper, we explore the relation between the density profile and the 
assembly histories of dark halos in CDM models. Our goal is to understand 
the physical processes that shape the density profiles of CDM halos, further 
motivated by the recent finding that the mass accretion histories of CDM halos 
show remarkable regularity. Based on high-resolution $N$-body simulations, 
the mass accretion history of a halo in general consists of two distinct
phases: an early fast phase and a later slow phase (Wechsler \etal 2002; 
Zhao \etal 2003a,b; Li et al.  2005). As shown in Zhao et al. (2003a), the 
fast accretion phase is dominated by major mergers characterised by a rapid 
deepening of the halo potential. The slow accretion phase is characterised by 
only weak changes in the gravitational potential well.  We will demonstrate
that the `universal' NFW profile results from gross properties of the mass 
accretion histories of dark matter halos.  Using a simple spherical collapse 
model, we show that the inner profile is established in the fast accretion phase.  
Its key features are rapid collapse (in a small fraction of a Hubble time)
with rapid changes in the potential that may isotropise the velocity field.
The outer profile is dominated by particles that are accreted in the 
slow-accretion phase onto an existing central object.  Our model incorporates 
these two phases of CDM accretion history to explain the appearance of 
`universal' density profile in $N$-body simulations.  In the absence of slow 
accretion, the outer profile would approach $\rho\propto r^{-4}$. Our findings 
suggest that mass accretion history plays a crucial role in structuring halos. 

The outline of this paper is as follows. In \S\ref{sec:mah}, we briefly 
describe CDM halo mass accretion histories, our procedure for realizing 
initial conditions for the formation of dark matter halos, and present our 
one-dimensional algorithm for simulating the collapse of dark matter halos. 
Our simulation results are described in \S\ref{sec:result}. In
\S\ref{sec:univ}, we use models with simple power-law initial perturbations 
to understand how NFW-like profiles are produced in CDM models. Finally, in 
\S\ref{sec:dis}, we further discuss and summarise our results.

\section{Initial conditions and methods} 
\label{sec:mah}

For a given halo, the mass accretion history specifies how much mass
is added to the halo as a function of time. Numerical simulations
and analytical models of halo formation in CDM models reveal that the
mass accretion histories of CDM halos are remarkably regular. Wechsler
\etal (2002) found that the accretion history of a CDM halos from
$N$-body simulations can be described roughly by the following
parametric form:
\begin{equation}
\label{eq_accretion}
M(a)=M_0 \exp\left[-a_c S\left({a_0\over a}-1\right)\right]\,,
\end{equation}
where $a$ is the expansion scale factor of the universe, and $M_0$ is
the virial mass of the halo at a final time where $a=a_0$. The
formation history is characterised by a single parameter $a_c$.  This
characteristic scale factor $a=a_c$ is the point when the logarithmic
mass accretion rate, ${{\rm d}\log M /{\rm d}\log a}$, falls below a
critical value $S=2$ (Wechsler et al. 2002). Similar results were
found by Zhao et al. (2003a; b) using high-resolution simulations and
by van den Bosch (2002) using extended Press-Schechter theory.

In a cosmological spherical collapse model, the shell collapse time is
determined by the mean initial over-density within the mass shell; the
mass within a mass shell collapses when the average linear
over-density (calculated using linear perturbation theory) reaches
$\delta_c\approx 1.686$.  Therefore, for a given mass accretion history, we can
then construct the corresponding initial density perturbation profile that
reproduces this history.
For a spherical perturbation of mass $M$ that collapses at a redshift
$z$, its linear over-density at the initial time $z_i$ is given by
\begin{equation}
\delta_i(M)= 1.686 \frac{D(z_i)}{D(z)}\,,
\label{eq_delta}
\end{equation}
where $D(z)$ is the linear growth factor.
In our calculation, we use the fitting formula by Carroll \etal (1992),
\begin{eqnarray}
D(z) &=& \frac{g(z)}{1+z} \\
g(z) &\approx & {5\over 2} \Omega_M(z) \nonumber
\left\{\Omega_M^{4/7}(z)-\Omega_{\Lambda}(z) \right.\\ 
&&\left.+\left[1+{\Omega_M(z)\over 2}\right] 
\left[1+{\Omega_{\Lambda}(z) \over 70}\right] \right\}^{-1},
\end{eqnarray}
where $\Omega_M(z)$ and $\Omega_\Lambda(z)$ are the density parameters
of non-relativistic matter and of the cosmological constant at redshift $z$,
respectively.  At redshift $z_i$, the radius $r_i$ of the sphere
that encloses mass $M$ is
\begin{equation}
r_i(M)=\left[\frac{3M}{4 \pi \bar{\rho}(z_i)
[1+\delta_i(M)]}\right]^{1/3}\,,
\label{eq_ri}
\end{equation}
where $\bar{\rho}_i(z_i)=\rho_{\rm crit,0} \Omega_M(0) (1+z_i)^3$, with
$\rho_{\rm crit,0}$ the critical density of the universe at $z=0$.
Assuming that $z_i$ is sufficiently large that the initial collapse is
linear, equation (\ref{eq_delta}) relates the enclosed mass $M(z)$ to
its overdensity.  Equation (\ref{eq_ri}) then determines its radius at
$z_i$.  Choosing a mass partition $M_1<M_2<\cdots<M_N$, one can then
determine the radii $r_1<r_2<\cdots<r_N$.  We choose equal mass shells
$m\equiv M_{j+1} - M_j$ for our simulations.  Note that this procedure
described by equations (\ref{eq_delta}) and (\ref{eq_ri}) is
independent of the mass scale $M_0$ in equation (\ref{eq_accretion}) and,
therefore, a single simulation may be scaled to any value of $M_0$ ex
post facto.

We use one-dimensional particle simulations to explore
the dynamical evolution of the collapse for a given mass accretion
history.  Note that even though the simulations are one-dimensional we can
include the effects of angular momentum (see eq. \ref{eq_j1d} below).
With a correction for the cosmological constant, we
approximate the gravitational acceleration of the $k$th mass shell
(particle) by
\begin{eqnarray}
g_k &=& H_0^2\Omega_{\Lambda}r_k-
\frac{GM_k r_k}{(r_k^2+{\alpha}^2)^{3/2}}\\
M_k &=& \sum_{r_j<r_k} m_j
\label{eq_cf}
\end{eqnarray}
where $H_0$ is Hubble's constant at the present time,
$\Omega_{\Lambda}$ is the density parameter of the cosmological
constant, $r_k$ is the radius of the shell, and $\alpha$ is a
softening length.  
In every simulation, we use a softening length $\alpha=0.0005r_v$, 
where $r_v$ is the virial radius of the halo at the present time. This
scale is much smaller than any scale of interest.
We assign each shell a specific
angular momentum $J_k$; $J_k\equiv0$ defines pure radial infall.
The effective acceleration including the centrifugal force is then 
\begin{equation}
a_k=g_k+{J_k^2 \over r_k^3} \label{eq_j1d}
\end{equation}

Since in spherical symmetry the gravitational force on a mass shell
is determined by the mass enclosed by the mass shell, we calculate the
force by sorting particles according to their radii.
Owing to the finite number of shells, when two shells cross they experience a
gravitational force discontinuity since then the enclosed mass instantly
changes.  To reduce such effects, we introduce another type of force softening.
Instead of assuming that each shell is infinitesimally thin, we assume that it
has a constant density with finite thickness, which we choose to be the
distance between the interior and exterior neighbouring shells.
Using such a softening method, crossing shells gradually change
their enclosed mass and hence the gravitational forces change smoothly.

We use a time-symmetric symplectic leapfrog integrator (Quinn \etal
1997; Springel 2005) to solve the equations of motion.  
At each time step, we calculate the dynamical time for every shell and
we choose the next time step to be smaller than the shortest dynamical
time of all the shells, i.e.
\begin{equation}
\Delta t_{dyn} = \min_k\left\{ {c_d\sqrt{5r_k^3 \over 2 GM_k}}\right\},
\end{equation}
where $c_d$ is control parameter, which we set to be 0.001.
Each simulation starts from an initial condition, which specifies the
position and velocity of each particle at a chosen high redshift. The
position for each particle at the initial redshift $z_i$ follows from
the mass accretion history as described above. The initial velocity
consists of two components: the Hubble expansion $v_i(M)=r_i(M)
H(z_i)$ and the peculiar velocity.  We use linear perturbation theory
to relate the initial peculiar velocity of a mass shell to
$\delta_i(M)$. We start our simulations at $z_i=200$, early
enough for linear theory to be valid.

All our simulations assume $\Omega_M=0.3$ and
$\Omega_{\Lambda}=0.7$. We use $10^5$ equal mass particles to simulate the
formation of a single dark matter halo, and we have tested that this
number is sufficiently large to achieve numerical convergence over the
scales in which we are interested.  In each simulation the total mass of
all the particles is 1.2 times the final virial mass of the
halo at $z=0$. The extra mass maintains an ambient environment to follow
the late stages of halo formation. As a test of our code we have reproduced
the self-similar results of both Fillmore \& Goldreich (1984) and 
Nusser (2001). We varied the particle number by a factor of 10, 
both larger and smaller, and could still satisfactorily reproduce 
these results.

\section{Results}\label{sec:result}

\begin{figure}
\centerline{\psfig{figure=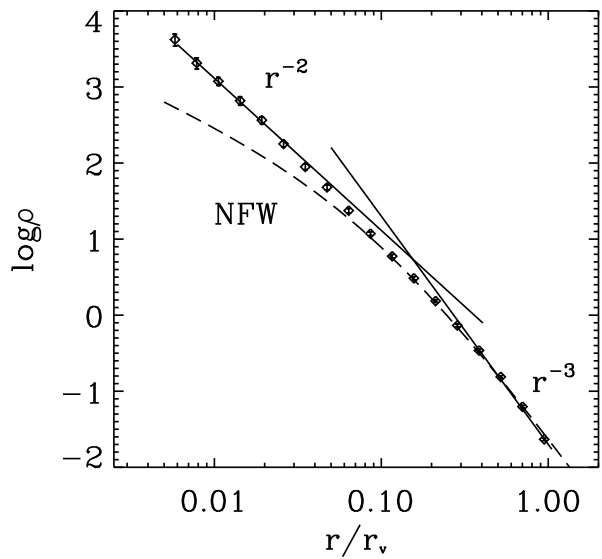,width=\hssize}}
\caption{The diamonds are the binned density profile of a simulated
dark matter halo at $z=0$ with Poisson error bars.  Particles are
assumed to have pure radial motion in this simulation. Note that the
outer density profile has $\rho\propto r^{-3}$, while the inner profile
has $\rho\propto r^{-2}$. The dashed curve shows a NFW profile for
comparison.}
\label{fig:prof1}
\end{figure}

We first consider a model with $a_c=0.4$ corresponding to a formation
redshift of $z_c=1.5$ with pure radial motion ($J_k\equiv0$ in
eq. \ref{eq_j1d}).  The density profile of the halo at $z=0$ is shown
in Figure \ref{fig:prof1}.  The final $z=0$ density profile has an
inner logarithmic slope of $-2$ and an outer slope of $-3$.  The
predicted inner profile is much steeper than a NFW profile, although
the model matches a NFW profile in the outer parts. We find that purely radial
collapse simulations with varying values of $a_c$ all produce halo profiles
with an inner logarithmic slopes of $-2$ and outer slopes of $-3$.
Differing values of $a_c$ affect only the transition radius between
these two slopes; larger values of $a_c$ (later formation times) leads to a
larger transition radius, i.e. a lower concentration.  
Radial infall model alone cannot reproduce a NFW
profile in the inner parts because it does not accurately represent
the dynamics of fast accretion.  The early fast accretion phase of a
halo is dominated by major mergers and the depth of the potential well
associated with the main progenitor deepens rapidly with time (Zhao
\etal 2003a). Frequent scattering by potential fluctuations in this phase 
is expected to effectively isotropise the 
orbits. As a result, dark matter particles will acquire a  
significant amount of angular motion as they are accreted by the halo, which
is not included in the purely radial calculation.
However, the radial infall model does appear to successfully describe
accretion from large distances onto an existing
central mass in the slow accretion regime of halo formation and we will
see in \S\ref{sec:univ} that this does explain the agreement of the model 
with a NFW density profile in the outer parts.
To simulate the fast accretion process more accurately using 
our one-dimensional approach, we consider a model that includes the 
effect of isotropic velocity dispersions. We trace 
the motion of each particle assuming pure radial motion into 
a radius of $R_t$, 
and at this point, we randomly assign a tangential component of 
velocity to the particle, keeping the kinetic energy unchanged.
As the simulation evolves further, the angular momentum of each
particle is conserved after an `isotropisation' event.  We choose $R_t$ to
be one half of the turn-around radius of each particle.  Although this
prescription seems somewhat arbitrary, a different choice only moderately
shifts the radial scale. In detail, we
incorporate this into our isotropisation prescription as follows.  For
a given mass shell $M$, we assume the ratio between the radial
velocity dispersion, $\sigma_r$, and the tangential velocity
dispersion $\sigma_t$, has the form,
\begin{equation}
\label{anisotropy}
{\sigma_t^2 \over \sigma_r^2} = 2 
\left[1+\left({R_{\rm t} \over r_a}\right)^{\beta}\right]^{-1}\,,
\end{equation}
where $r_a$ is the characteristic scale of the halo demarcating fast
and slow accretion and $\beta$ controls the shape of the anisotropy
profile. 
The radial and tangential velocities are
randomly sampled from Gaussian distributions. 
The ratio of these two random numbers are then used to 
partition the kinetic energy of the particle into radial and
tangential components.
For mass shells with
$R_{\rm t}\ll r_a$, the velocity distribution is nearly isotropic,
while for those with $R_{\rm t}\gg r_a$ radial orbits dominate.  Note
that a larger $\beta$ produces a sharper transition between radial and
isotropic orbits. We take $\beta=2$ as our fiducial model and will
describe the effect of changing the value of $\beta$ below.

\begin{figure}
\centerline{\psfig{figure=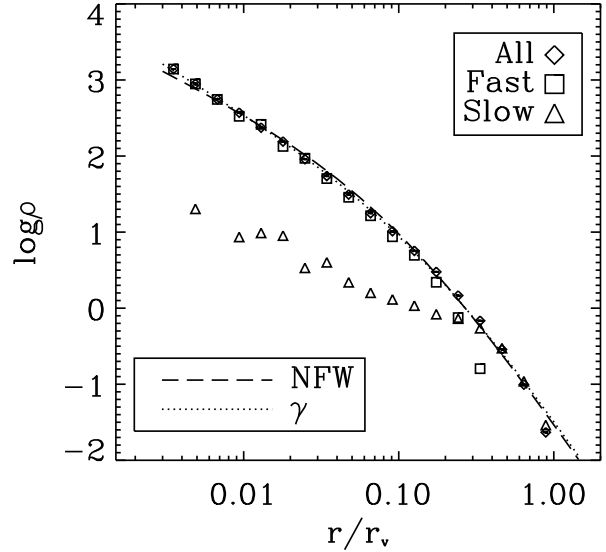,width=\hssize}}
\caption{The density profile of a dark matter halo (diamonds) in a
simulation where particles accreted in the fast accretion phase are
assumed to have isotropic velocity dispersion (see text for
details). 
The simulated density profile can be fit by a NFW profile (dashed curve). 
The dotted curve is the result of the best fit to the data
by equation (\ref{rho_gamma}), and the best-fit value for $\gamma$ is $1.21$. 
Squares show the density
profile for particles that are accreted in the fast accretion phase,
while triangles show the density profile for particles accreted in the
slow accretion phase.  Note that the outer part of the halo profile is
dominated by particles in slow accretion, while the inner profile is
dominated by particles in the fast accretion phase.}
\label{fig:prof2}
\end{figure}

Since $R_{\rm t}$ is roughly the virial radius of a mass shell, we
take $r_a=r_v(a_c)$, where $r_v(a_c)$ is the virial radius of the halo
at the transition time between the fast accretion phase and the slow
accretion phase.
Figure \ref{fig:prof2} shows the final halo density profile of such a
simulation.  The resulting inner density profile is much flatter than
the $\rho\propto r^{-2}$ we find for pure radial motions, and now
over a large range
of radius it can be well fit by a NFW profile.  The inner profile
is dominated by particles accreted in the fast accretion regime, and
the shallower profile owes to the non-radial motions of these
particles.  These results are not particular to the mass accretion
history used in this example, and indeed our simulations with other
mass accretion histories (see below) all lead to similar results.
 
\begin{figure*}
\centerline{\psfig{figure=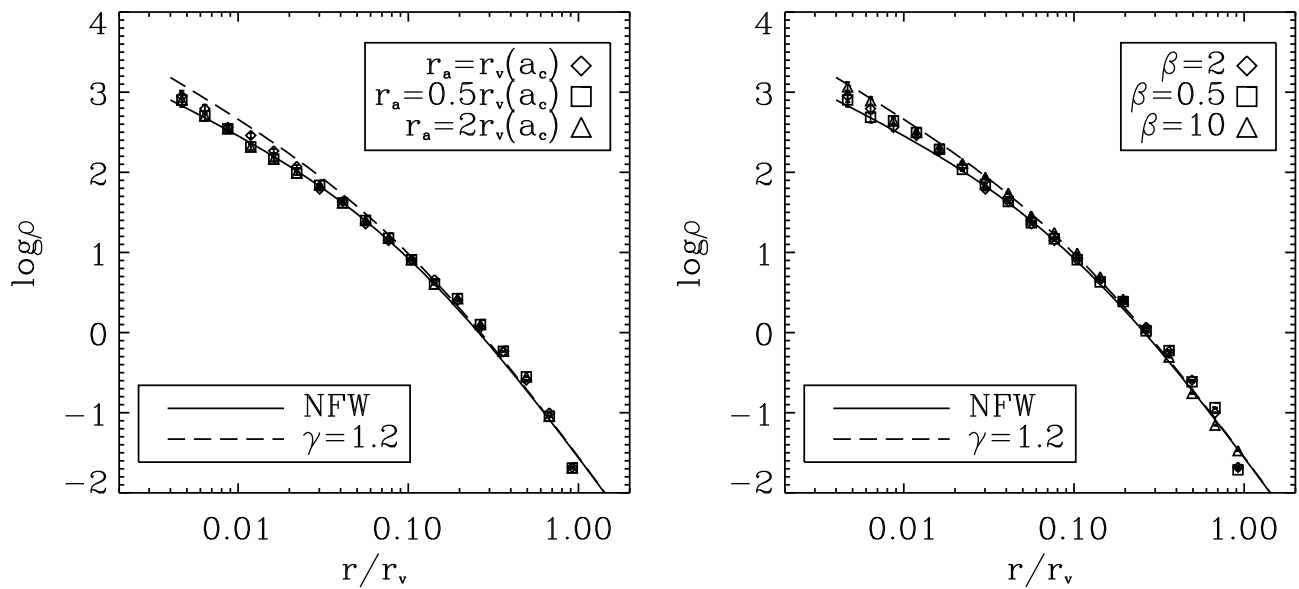,width=\hdsize}}
\caption{The density profiles for models with different choices of
$r_a$ and $\beta$ [see equation (\ref{anisotropy}) for definition].
In the left panel, results are shown for models where $\beta$ is fixed
to be $2$ but $r_a$ changes from $r_v(a_c)$ (diamonds) to
$0.5r_v(a_c)$ (squares) and $2r_v(a_c)$ (triangles). In the right
panel, $r_a$ is fixed to be $r_v(a_c)$ but the value of $\beta$
changes from $2$ (diamonds) to $0.5$ (squares) and $10$
(triangles). The solid curve shows a NFW profile, while the dashed
curve shows the profile (eqn. \ref{rho_gamma}) with $\gamma =1.2$.  }
\label{fig:prof3}
\end{figure*}

We check the sensitivity to our velocity structure assumptions by
varying the values of the two parameters $r_a$ and $\beta$ in equation
(\ref{anisotropy}). We first consider two alternatives to the fiducial
model $r_a= r_v(a_c)$: $r_a=0.5 r_v(a_c)$ and $2r_v(a_c)$ with fixed
$\beta=2$.  The density profiles given by these two models are shown
in the left panel of Figure \ref{fig:prof3}.  The density profile
changes very little even though the value of $r_a$ changes by a factor
of 4.  Next, we hold $r_a$ fixed at our fiducial value of $r_v(a_c)$ and
vary the value of $\beta$ from $0.5$ to $10$. The resulting density
profiles are shown in the right panel of Figure \ref{fig:prof3}.  The
final profile is also insensitive to changes in $\beta$.

\emph{In summary, if particles accreted in the fast accretion regime
are assumed to have an isotropic velocity dispersion, we find that
the inner slope of the halo density profile always lies in the range
between $-1$ and $-1.2$.  Apparently, the inner $\rho\propto r^{-1}$
profile is a generic result of the fast collapse phase and the
isotropic velocity field generated during such a collapse.}

The results obtained above can be compared to those obtained earlier
by Avila-Reese \etal (1998) and Ascasibar \etal
(2004). These authors examined the density profiles of dark matter
halos produced by spherical collapse in a CDM density field, assuming
that the orbits retain a constant ellipticity.  These models can produce
NFW-like profiles provided that the constant ellipticity is properly
chosen.  However, as we describe in \S\ref{sec:univ}, spherical
collapse with constant ellipticity orbits can produce a wide range
of inner profiles depending on initial conditions.  A good match
between their models and the NFW profile requires a fine tuning of
initial conditions. Shapiro et al. (2004) also explored the
importance of CDM accretion history using one-dimensional simulations
and very similar arguments to ours.  Unfortunately, they use a fluid approach
that solves Jeans-like moments of the collisionless Boltzmann equation and,
presumably, the elimination of any possible asymmetric velocity 
distribution prevented them from finding our $\rho\propto r^{-1}$ result.

In contrast to these models, our consideration is based on realistic CDM 
halo formation histories. We demonstrate that, for all such formation 
histories, the early fast
accretion of dark matter that may effectively generate an isotropic 
velocity dispersion leads naturally to $\rho\propto r^{-1}$ in the inner
parts.
Our isotropisation assumption is supported by the measurement of 
halo velocity dispersions in cosmological $N$-body simulations, 
which becomes progressively more isotropic
towards the inner part of dark halos (e.g. Eke \etal 1998; Col\'{i}n \etal
2000; Fukushige \& Makino 2001; Hansen \& Moore 2004).
 
\begin{figure}
\centerline{\psfig{figure=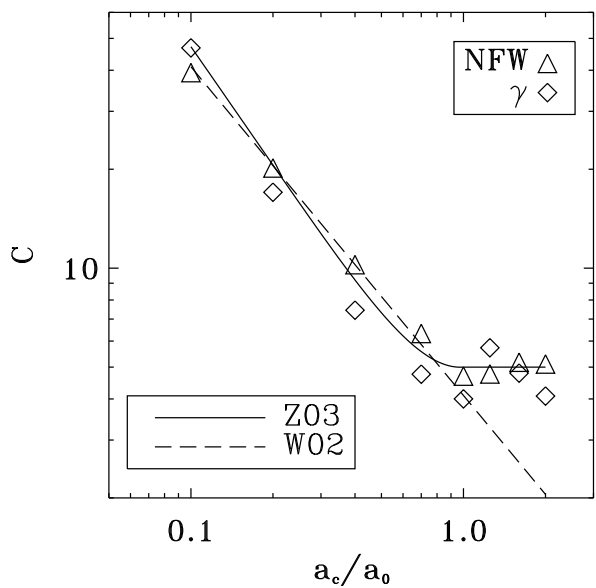,width=\hssize}}
\caption{Halo concentration as a function of $a_c$, which characterises
the halo formation time. Triangles show the results where
simulated halo profiles are fit with a NFW profile and the
diamonds are the results when fit with
equation (\ref{rho_gamma}). The solid curve shows the prediction of a
model proposed by Zhao \etal (2003a) based on cosmological $N$-body
simulations, while the dashed curve shows the model prediction given
by Wechsler \etal (2002).}
\label{fig:ca}
\end{figure}
 
We can study the dependence of halo
structure on the `formation time', characterised by $a_c$, using our
one-dimensional simulations. We fit the resulting density
profiles both with a NFW profile and one with the following more general form:
\begin{equation}
\label{rho_gamma}
\rho(r)={\rho_s \over 
(r/r_s)^{\gamma}\left[1+(r/r_s)\right]^{(3-\gamma)}}\,,
\end{equation}
where $\gamma$ is the inner slope of the profile.  Figure \ref{fig:ca}
shows the best fit value of the concentration parameter, defined as
$c\equiv {r_v/r_s}$, as a function of $a_c$. As one can see, halos
with lower $a_c$, i.e. earlier formation times,  are more concentrated.
However, for halos that are still in the fast accretion phase, i.e.  with 
$a_c>1$, the concentration is independent of $a_c$.  The solid curve shows the
concentration-formation time relation obtained by Zhao \etal (2003a;b)
using high-resolution $N$-body simulations. Our results assuming a NFW
profile match this relation extremely well. The dashed line shows the
model proposed by Wechsler \etal (2002). This model agrees well with
our results for $a_c<1$, but underestimates the concentration for
halos with $a_c>1$. We refer readers to Zhao et al. (2003a;b) for a
detailed discussion about the discrepancy between their results
and those obtained by Wechsler \etal (2002).
Since it is known that the concentration-formation time relation is the
origin of the mass-concentration relation (Wechsler \etal 2002,
Zhao \etal 2003a), our results will also reproduce the mass-concentration
relation (at a given redshift) as well as the concentration-redshift 
relation (at fixed mass) found in the cosmological $N$-body simulations.

\begin{figure*}
\centerline{\psfig{figure=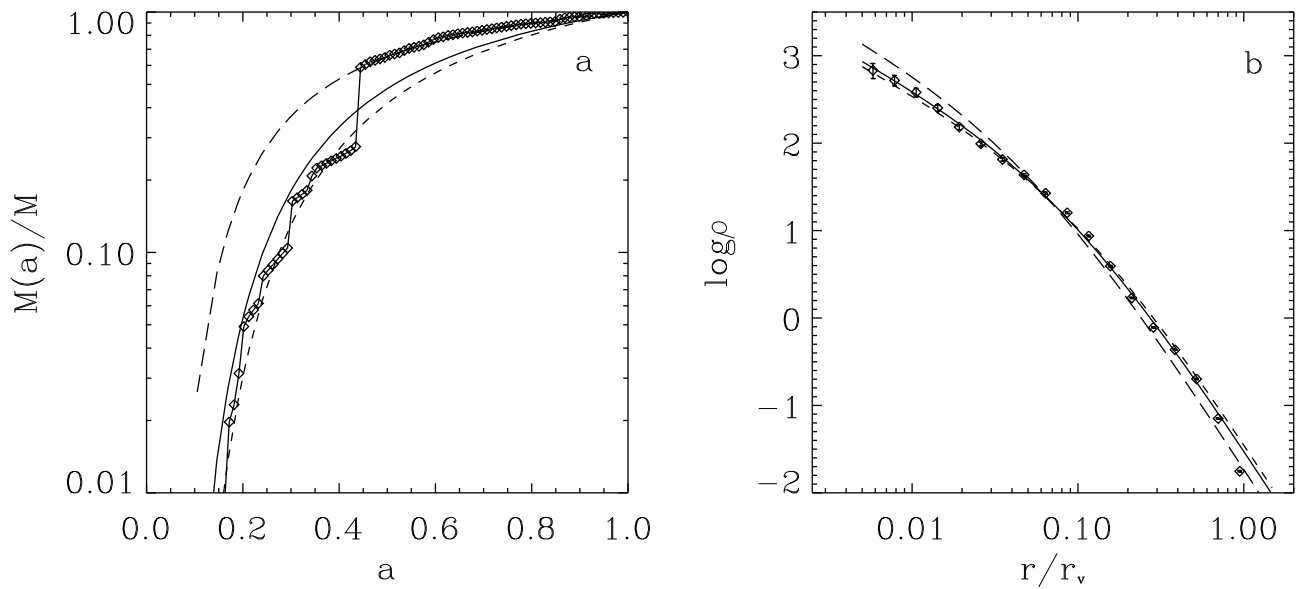,width=\hdsize}}
\caption{The diamonds in panel (a) show a mass accretion history 
that cannot be well fit using
equation (\ref{eq_accretion}). The long (short) dashed curve 
is the result of fit that emphasises the recent (past) history
at $a>0.45$ ($a<0.45$). The solid curve represents a 
compromise between the recent and past histories. 
The halo profiles corresponding to the actual mass accretion 
history and these three possible fits are shown in panel (b).
}
\label{fig:special}
\end{figure*}

So far, our discussion is based on simulations that assume the smooth
accretion history given by equation (\ref{eq_accretion}).  Although
this smoothed form is a good description of the accretion history
averaged over an ensemble of simulated dark halos, an individual
accretion history exhibits details that are not described by equation
(\ref{eq_accretion}).  In some cases, equation (\ref{eq_accretion})
even fails to describe the overall shape of the mass accretion history.
We show such an example in the left panel of Figure \ref{fig:special}.
A fit that emphasises the accretion history before the big jump at $a=0.45$,
shown by the short dashed curve, differs from a fit that emphasises the
history after that time.  To see how such differences affect
the density profiles, we carried out one-dimensional
simulations using realistic halo accretion histories generated by
PINOCCHIO, a Lagrangian code developed by Monaco et al. (2002).  The
statistical properties of the mass accretion histories generated using
this code agrees with those obtained using cosmological $N$-body simulations
(Li et al. 2005).  The resulting simulated halo density profiles can all be
well described by the NFW form. For many cases where equation
(\ref{eq_accretion}) is a good fit to the overall accretion history,
the simulated density profiles using real accretion histories are very
similar to those using the corresponding fits with equation
(\ref{eq_accretion}).  Even for cases where the accretion history is
poorly described by equation (\ref{eq_accretion}), as the one shown in
Figure \ref{fig:special}a, the final halo profile can still be well fit
using a NFW form (see Fig.\,\ref{fig:special}b).  In such cases,
however, the formation time defined using equation (\ref{eq_accretion})
is uncertain.

\begin{figure}
\centerline{\psfig{figure=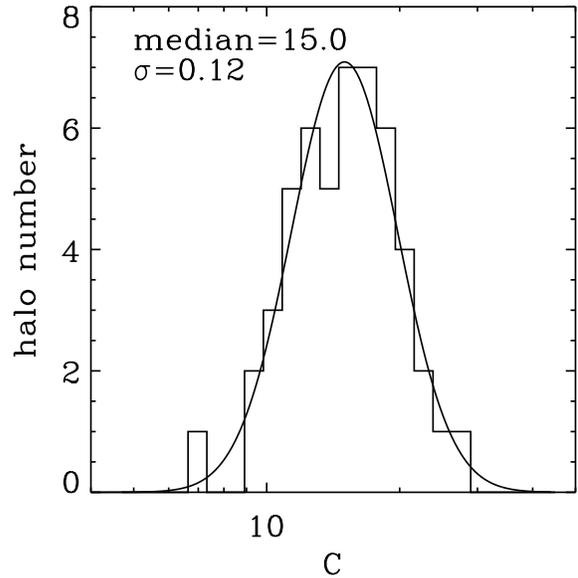,width=\hssize}}
\caption{
The histogram shows the distribution of halo concentration 
obtained from an ensemble of 50 randomly chosen mass accretion histories
for halos with masses in the range $10^{11}\Msun$ to 
$10^{12}\Msun$ at the present time. The solid curve
shows a log-normal distribution with a median 
equal to $15.0$ and a dispersion (in $\log c$) 
equal to $0.12$. 
}
\label{fig:cdist}
\end{figure}

The difference in the detailed mass accretion history
introduces a scatter in the distribution of the concentration parameter
$c$ (Wechsler \etal 2002). If the density profile is determined by its
mass accretion history, then we should be able to reproduce the
distribution of $c$ using a random sample of realistic mass accretion
histories.  To perform this experiment, we have randomly chosen 50
mass accretion histories generated with PINOCCHIO for halos with
masses in the range $10^{11}-10^{12}\Msun$. These mass accretion
histories are used to generate initial conditions for our
one-dimensional simulations. Figure \ref{fig:cdist} shows the
distribution of the concentration parameter, obtained by fitting the
simulated profiles with the NFW form. The distribution can be roughly
described by a log-normal, in agreement with the cosmological
$N$-body results (e.g. Jing 2000; Bullock \etal 2001).  The
dispersion in $\log c$, $\sigma=0.12$, is also very close to that
found in cosmological $N$-body simulations (e.g. Wechsler \etal 2002),
reinforcing our finding that the density profile of a halo is 
largely determined by its mass accretion history.

\section{What determines the density profiles of dark matter halos?}
\label{sec:univ}

In the last section, we have shown that many of the structural properties of
CDM halos found in cosmological $N$-body simulations
can be understood in terms of halo mass accretion histories.  However,
why does gravitational collapse with such initial conditions always produce
halo profiles that follow a universal form? Is the universal profile a
result of the fact that the initial conditions represented by CDM mass
accretion histories are special, or is it more generic in the sense that it
can be produced by a wide range of initial conditions?

To answer this question, we consider the collapse of generic
initial perturbations with the form 
\begin{equation}
\label{pert}
{\delta M(r) \over M(r)}\propto M(r)^{-\epsilon}\,,
\end{equation}
where $\epsilon$ is a constant that controls the mass accretion
rate. Because a mass shell collapses when $\delta M/M\approx 1.68$, the
mass accretion history implied by equation (\ref{pert}) is $M(z)
\propto D^{1/\epsilon}(z)$, where $D(z)$ is the linear growth
factor. Since the circular velocity of a halo $V_c$ is related to its mass,
$M\sim V_c^3/H(z)$, where $H(z)$ is the Hubble constant at redshift $z$,
we can write $V_c(z) \propto H^{1/3}(z)
D^{1/3\epsilon}(z)$.  For simplicity, we consider an Einstein-de
Sitter universe.  In this case, $D\propto H^{-2/3}$, and therefore
\begin{equation}
M\propto H^{-2/3\epsilon}\,,
~~~~
V_c\propto H^{(1-2/3\epsilon)/3}\,.
\end{equation}  
Note that $V_c^2$ is a measure of the depth of potential well associated
with the halo.
For $\epsilon=1/6$, $V_c\propto H^{-1}$ ($M\propto H^{-4}$).
Therefore, $\epsilon=1/6$ separates the isotropisation
regime from calm accretion, i.e. $V_c$ changes by an order of
unity or more in a Hubble time for $\epsilon<1/6$.  For
$\epsilon=2/3$, $M\propto H^{-1}$ ($V_c={\rm constant}$) and, similarly,
$\epsilon=2/3$ separates fast accretion from slow accretion in terms
of the mass accretion rate.  For $\epsilon \to 0$, perturbations on
different mass scales have the same amplitude, and all mass scales
collapse simultaneously, leading to fast increases in both mass
and potential well depth.  In contrast, for $\epsilon
\gg 1$ the perturbation amplitude declines rapidly with mass scale, hence
$M$ increases little while the potential well decays as the
universe expands.  This allows us to study cases with vastly different collapse
histories by changing the value of $\epsilon$ over a large range.

\begin{figure}
\centerline{\psfig{figure=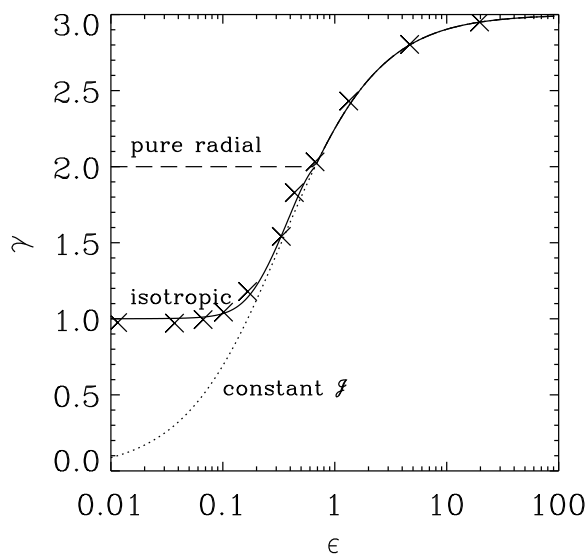,width=\hssize}}
\caption{The relation between the inner logarithmic slope,
$\gamma$ ($\rho\propto r^{-\gamma}$),
as a function of the exponent of the initial perturbation
defined in equation (\ref{pert}).  The long dashed curve shows the
solution of the model with pure radial motion 
and the dotted curve shows the solution when
all particles have the same ${\mathcal J}$.  The
solution with an isotropic velocity dispersion is shown as
the solid curve, and is compared with the result obtained directly
from numerical calculations (crosses). All the three curves overlap
for $\epsilon\ge 2/3$.}
\label{fig:g_e}
\end{figure}

In an Einstein-de Sitter universe where the expansion factor $a$ is a
power law of time, the collapse of perturbations described by
equation (\ref{pert}) admits similarity solutions.  Assuming spherical
symmetry and pure radial motion, Fillmore \& Goldreich (1984) show
that the collapse develops an asymptotic density profile $\rho \propto
r^{-\gamma}$ in the inner region, with $\gamma=2$ for $0 < \epsilon
\leq 2/3$, and $\gamma=9\epsilon/(1+3\epsilon)$ for $\epsilon >
2/3$. These solutions are shown in Figure \ref{fig:g_e} as the
long-dashed curve. Self-similar models can also be constructed for
non-radial motions (Nusser 2001). The specific angular momentum of a
particle can be specified as
\begin{equation}
J={\mathcal{J}}\sqrt{G M_{\rm ta} r_{\rm ta}}\,,
\end{equation} 
where $r_{\rm ta}$ is the turnaround radius of the particle, and
$M_{\rm ta}$ is the mass interior to $r_{\rm ta}$. If ${\mathcal{J}}$
is the same constant for all particles, the problem admits similarity
solutions and the asymptotic inner slope of the density profile is
$\gamma={9\epsilon/(1+3\epsilon)}$ for all $\epsilon >0$ (Nusser
2001). This solution is shown as the dotted curve in
Figure \ref{fig:g_e}. For $\epsilon\ge 2/3$, this solution is the same
as that for the model with pure radial infall.  The asymptotic value
of $\gamma$ is $0$ as $\epsilon\to 0$.  The dependence of $\gamma$ on
$\epsilon$ for a constant ${\mathcal{J}}$ is steep near $\gamma\sim
1$. Thus, to produce a NFW inner slope in this model requires
tuned initial conditions.

For our models in \S\ref{sec:result}, the velocity dispersion of
particles in the early collapse phase is isotropic. Thus, the quantity
${\mathcal J}$ has the following distribution at fixed energy:
\begin{equation}\label{eq:L_dist}
P({\mathcal J})={{\mathcal J} \over \sqrt{1-{\mathcal J}^2}}\,.
\end{equation}
For the slow accretion phase, we expect the results of pure radial
motion to obtain.  Thus, for $\epsilon \ge 2/3$, the asymptotic slope
is $\gamma={9\epsilon/(1+3\epsilon)}$.  For $\epsilon < 2/3$, we can
also use a simple model to understand the results obtained in the last
section.  Consider all particles with a given ${\cal J}$.  In general,
we can write the final density profile of these particles as
$\Delta\rho_{\cal J}(r) \propto(\Delta M_{\cal J}/ r_{\cal J}^3)
F(r/r_{\cal J})$, where $\Delta M_{\cal J}$ is the total mass of
particles with ${\cal J}$ in the range ${\cal J}\pm\Delta {\cal J}$,
and $r_{\cal J}$ is a characteristic scale in the density profile.
The quantity $\Delta M_J/r_J^3$ is the density scale.  Since the only
scale in the problem is the current turnaround radius $r_{\rm out}$,
and since non-radial motion is expected to be important only for $r\ll
{\cal J} r_{\rm out}$, we expect $r_{\cal J}\propto {\cal J} r_{\rm
out}$.  Thus,
\begin{equation}
\Delta\rho_{\cal J}(r) \propto 
{M\over r_{\rm out}^3}
{P({\cal J})\Delta {\cal J}
\over {\cal J}^3} F\left({r\over {\cal J}r_{\rm out}}\right)\,.
\end{equation}
If we neglect the interaction between mass shells of 
different ${\cal J}$, the total density can be written as
\begin{equation}
\label{rhoasJ}
\rho(r)=\sum\Delta\rho_{\cal J}(r)
\propto {M\over r_{\rm out}^3} 
\int_0^1 {P({\cal J}){\rm d} {\cal J}\over {\cal J}^3}
F\left({r\over {\cal J}r_{\rm out}}\right)\,.
\end{equation}

On scales where the effect of angular momentum is important,
i.e. $r\ll {\cal J} r_{\rm out}$, the density profile is expected to
be $F(r)\propto r^{-9\epsilon/(1+3\epsilon)}$.  Conversely, for $r\gg
{\cal J} r_{\rm out}$ the particles are in nearly radial
motion and one expects $F(r)\propto r^{-2}$ (for $\epsilon <2/3$).
Hence, we may approximate the form of $F$ as
$F(x)=1/[x^2+x^{9\epsilon/(1+3\epsilon)}]$. Inserting this and
equation (\ref{eq:L_dist}) into equation (\ref{rhoasJ}), we obtain
\begin{equation}
\rho(r)
\propto {M\over r_{\rm out}^3}
{r_{\rm out}\over r}
Q(\epsilon, r_{\rm out}/r)\,,
\end{equation}
where 
\begin{equation}
Q\left(\epsilon,r_{\rm out}/r \right)
=\int_0^{\pi/2}{(r_{\rm out}/r){\rm d}\theta\over 
1+[(r_{\rm out}/r)\sin\theta]^\alpha}\,,
\end{equation}
with 
\begin{equation}
\alpha\equiv {2-3\epsilon \over 1+3\epsilon}\,.
\end{equation}
The inner density profile is given by the $r$ dependence of $Q$ for
$r/r_{\rm out}\to 0$. The solid curve in Figure \ref{fig:g_e} shows the
resulting $\gamma$-$\epsilon$ relation. The characteristics of this
relation can be understood as follows.  For $\epsilon <1/6$, then
$\alpha>1$.  The integration in $Q$ is dominated by small $\theta$,
and so we can replace $\sin\theta$ by $\theta$.  The function $Q$ is
independent of $r$ and $\rho\propto r^{-1}$. For
$1/6<\epsilon <2/3$, then $0<\alpha<1$ and the integration is now dominated
by $\sin\theta>r/r_{\rm out}$.  The function $Q\propto r^{\alpha-1}$
and, therefore, $\gamma={9\epsilon/(1+3\epsilon)}$.  This relation between
$\gamma$ and $\epsilon$ holds also for $\epsilon >2/3$, as discussed
above. 

To check the accuracy of the above simple model, we use a
numerical calculation to solve for the inner density profile as a function
of $\epsilon$. For any radius $r$, the total mass within it can be
written in two parts,
\begin{equation}
m_T(r)=m_p(r)+m_t(r)\,.
\end{equation}
Here $m_p$ is the mass of all particles with apocenter smaller than
$r$. We call these particles `permanent' contributors because they
always contribute to $m_T(r)$. The mass $m_t$ in the above equation is the
contribution of particles with apocenter $r_a$ larger than $r$ but with a smaller
pericenter $r_b$. These are `temporary' contributors because
they only spend part of their orbital times within $r$. Let $P(r\vert
r_k)$ be the fraction of time that a `temporary' particle with
turnaround radius $r_k$ spends inside radius $r$.  The total mass
contributed by temporary particles can be written as
\begin{equation}
\label{eq:m_t}
m_t(r)=\int_{r_b<r<r_a} 
P(r\vert r_k) {\rm d} m(r_k),
\end{equation}
where the measure includes all orbits that pass through the surface at
$r$.
By definition,
\begin{equation}
\label{eq:P}
P(r\vert r_k)={\int_r ^{r_a} {{\rm d}r' \over 
v_k(r')} 
\Big/ \int_{r_b} ^{r_a} {{\rm d}r' \over v_k(r')}}
\end{equation}
where $v_k(r')$ is the radial velocity of a particle with turn-around
radius $r_k$ at a radius $r'$. Note that $v_k(r)$ as a function of $r$
depends on the current mass profile $m_{T}(r)$, and so the mass
profile has to be solved by iteration. The radial velocity can be
written as
\begin{equation}
v_k(r)=\sqrt{2}\left[E_k-\Phi(r)-{J_k^2/(2r^2)}\right]^{1/2}
\end{equation}
where $E_k$ and $\Phi$ are, respectively, the total energy and
gravitational potential of the particle, and $J_k$ is the specific
angular momentum. The increase of mass within the apocenter of a
particle can change the orbit of the particle.  Thus, we must
recompute such changes at each step of the iteration.  Assuming that
angular momentum is conserved, the apocenter $r_a$ of a particle after
a given iteration is related to the apocenter $r_a'$ before
by $m_{T}(r_{a}) r_a=m'_{T}(r_a') r_a'$, where $m'_{T}$ and
$m_{T}$ are the mass profiles before and after the iteration 
step. The iteration starts from an initial condition where each
particle is at its turnaround radius, and we denote the corresponding
profile by $M_{\rm ta}(r_{\rm ta})$.  Since the turnaround radius
$r_{\rm ta}$ of a mass shell is related to its initial radius $r_i$ by
$r_{\rm ta}\propto r_i (\delta M/M)^{-1}$ in an Einstein-de Sitter
universe and since $M\propto r_i^3$, one can show that
\begin{eqnarray}
M_{\rm ta}(r_{\rm ta}) \propto r_{\rm ta}^n\,, ~~~~ 
n={3/ (3\epsilon +1)}\,.
\end{eqnarray}

We have applied the above numerical calculation to models with pure
radial infall and models where ${\cal J}$ is the same for all mass
shells. The results match those given by the self-similarity
solutions.  The crosses in Figure \ref{fig:g_e} show the
$\gamma$--$\epsilon$ results for this calculation using an isotropic
velocity dispersion given by equation (\ref{eq:L_dist}).  The
numerical results match the simple analytical model presented
above remarkably well. We have also applied our one-dimensional code
to this model and the results are similar to those
obtained here.

One important feature in the $\gamma$-$\epsilon$ relation predicted by
this model is that $\gamma \approx 1$ for all $0< \epsilon\la
1/6$. For a perturbation with $\epsilon$ in this range, the circular
velocity of the halo increases rapidly with time, with a timescale
that is shorter than a Hubble time. In this case, not only can particles
with small apocenters (low orbital energies) reach the inner part
of the halo, but also can many particles with large apocenters (high
orbital energies) and small angular momenta. Velocity isotropisation 
mixes these orbits, resulting in $\gamma\approx 1$ as we have 
demonstrated. For $\epsilon>1/6$
the gravitational potential is changing gradually and particles joining
the halo have a similar energy and orbital shape ${\mathcal J}$.  The
resulting profile can then be described by the self-similar solution that
assumes the same orbital shape for all particles and $\gamma$ becomes larger
than one.  Note that an inner
logarithmic slope of approximately $-1$ results from a fast collapse
and orbit isotropisation and that both conditions are required to
produce such an inner slope. If the collapse is fast ($\epsilon \la
1/6$) but the velocity dispersion is not isotropic, the inner slope
can be as shallow as 0 (for constant ${\cal J}$) and as steep as $-2$
(for radial infall).  If the velocity dispersion is isotropic, but the
mass accretion rate is small, i.e. $\epsilon>1$, the inner slope can be
much steeper than $-1$.

The identification of this mechanism not only explains the approximate
universality of the inner CDM halo slopes but also describes the
variance observed in $N$-body simulations.  Gravitational collapse
starts on small scales in a hierarchical model such as CDM.  Deep
dark-matter potential wells are created by subsequent non-linear
gravitational collapse,  but the potential well associated with a halo
cannot deepen significantly during the slow accretion phase when the
accretion time scale is longer than a Hubble time (Zhao et
al. 2003a).  Therefore, all halos must have gone through a phase of
rapid accretion to establish their potential wells, even though
different halos may have different mass accretion histories.
Also, as shown in Li et al. (2005), the
phase of rapid mass accretion is dominated by major mergers, which may
effectively isotropise the velocity field.  These are the two
ingredients that are required to produce a $\rho \propto r^{-1}$ 
inner profile, and explain why such a inner profile results for
halos with vastly different formation histories. If the potential well
of a halo is established at early times, the mass contained in the
$r^{-1}$ profile will be small, and the halo will have a high concentration
since most of the mass accreted slowly. However,
if a halo establishes its potential well recently, much of
its mass will be in the $r^{-1}$ profile, and the halo will have a low
concentration. This correlation between halo concentration and the time of
potential well formation matches cosmological $N$-body simulations
(e.g. Zhao et al. 2003a;b).  In extreme cases where the mass involved
in the fast accretion is too small to be seen, we expect an inner profile
steeper than $r^{-1}$.

So far we have only considered the origin of the inner $r^{-1}$
profile.  What about the outer $r^{-3}$ form in the universal profile?
Consider a shell of mass $M$ and of an initial radius
$r_i$. Suppose this mass shell collapses at a time $t$ to a radius
$r$.  Assuming an Einstein-de Sitter Universe and neglecting the
effect of shell crossing, we have $r\propto r_i/\delta_i(M)$ and
$t\propto t_i/\delta_i(M)^{3/2}$. Eliminating $\delta_i (M)$ in these
two relations and using $M\propto r_i^3$ we obtain $r\propto M^{1/3}
t^{2/3}$. Thus, the density profile can be written as
\begin{equation}
\rho(r)={1\over 4\pi r^2}{{\rm d} M\over {\rm d}t} 
   \left({{\rm d} r\over {\rm d} t}\right)^{-1}
\propto
{M\over r^3}
{\mu \over 2+\mu}\,,
\end{equation}
where $\mu\equiv {\rm d}\ln M/{\rm d}\ln t$. Now let us write the mass
in two parts, $M=M_e+\Delta M$, where $M_e$ is the mass of the halo at
$t-\Delta t$, while $\Delta M$ is the mass accreted between time
$t-\Delta t$ and $t$. If $\Delta M$ increases as a power law of $t$,
and if $\Delta M\ll M_e$, then $\rho\propto r^{-3}$.  Since a mass
shell settles into an equilibrium state over several dynamical times,
the relevant scale for $\Delta t$ is the dynamical time of the
system. Thus, if there is a period of slow accretion,
but where the total accreted mass is much smaller than
the mass that has collapsed into the halo, an outer density
profile with $\sim r^{-3}$ results. However, the above
argument also suggests that the outer density profile can change from
halo to halo, depending on the mass accretion rate at the final stage
of halo formation. For example, if $M$ continues to grow as described by
equation (\ref{eq_accretion}) eventually a
density profile with $\rho\propto r^{-4}$ in the outer parts will result.

\begin{figure*}
\centerline{\psfig{figure=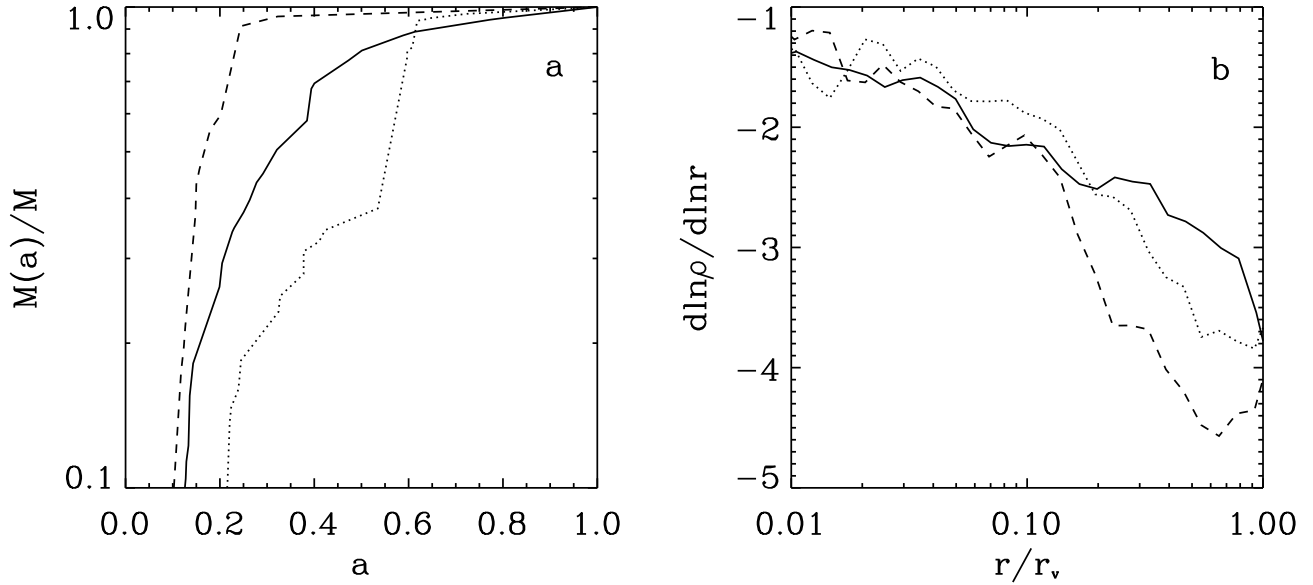,width=\hdsize}}
\caption{The left panel shows three mass accretion 
histories that have different accretion rates in the recent 
past. The right panel shows, as a function of radius,  
the logarithmic slope of the density profile generated with 
each of these three mass accretion histories.
Note that the outer slope can be as steep as $-4$ 
for a halo whose recent mass accretion rate is very small
(dotted and dashed curves). 
}
\label{fig:outprof}
\end{figure*}

To demonstrate such a dependence of the outer profile on mass
accretion history, we have carried out 1-D simulations for three cases
with different recent mass accretion histories. These mass accretion
histories are shown in the left panel of Figure \ref{fig:outprof}, and
the resulting logarithmic slopes versus $r$ are shown in the right
panel.  This figure shows that while halos with typical mass accretion
histories have $\rho\propto r^{-3}$ outer density profiles, halos with
slower accretion rates in the recent past (shown by the dashed and
dotted curves) have steeper outer profiles.  Therefore, the outer
$r^{-3}$ profile is not universal but a consequence of the form of
mass accretion history of typical CDM halos and in the currently accepted
cosmological model all halos will have a $r^{-4}$ outer profile
in the distant future.

\section{Summary and Discussion}\label{sec:dis}

We have constructed a simple model that reproduces many of the
properties of the CDM halo population.  There are two essential
ingredients: 1) a two-phase accretion history beginning with rapid
accretion and potential-well deepening followed by slow accretion with
little change in the potential well; and 2) isotropisation of orbits
during the phase of rapid growth.  The density profiles obtained from
various mass accretion histories all fit the NFW form. Our model
also reproduces the correlation between the concentration and
formation time observed in cosmological $N$-body simulations.  In
particular, the model predicts a roughly constant concentration,
$c\sim 5$, for all halos that are still in the fast accretion phase,
matching the results obtained by Zhao et al. (2003a;b). Combined
with an ensemble of realistic mass accretion histories parametrised
from cosmological CDM simulations, the model reproduces the dependence
of halo concentration on halo mass and the distribution of halo
concentrations measured in these same cosmological $N$-body
simulations.  Our results demonstrate that the structural properties
of CDM halos are largely determined by their mass accretion histories.

We can recover many of these results using a simple analytic model.
Our model begins with scale-free perturbations of the form $\delta M(r)/M(r)
\propto M(r)^{-\epsilon}$ in an Einstein-de Sitter universe. Assuming an
isotropic velocity dispersion, we find that the inner profile
produced by the collapse of such perturbations is always a power law,
$\rho(r)\propto r^{-\gamma}$, with $\gamma=1$ for $0<\epsilon < 1/6$,
and $\gamma=9\epsilon/(1+3\epsilon)$ for $\epsilon>1/6$.  A model with
$0<\epsilon < 1/6$ has a rapidly deepening potential leading to orbit
isotropy.  This produces a shallower profile than a purely radial model.
Assuming non-radial orbits of constant shape yields self-similar models but
with a large range of possible inner slopes.  The mixture resulting
from isotropisation converges to a single inner slope $\rho\propto r^{-1}$.
This suggests that the inner $r^{-1}$ profile of CDM halos is a
natural result of hierarchical models, where the potential well
associated with a halo has to be built through a phase of rapid and
violent accretion in which potential fluctuations are expected to 
effectively isotropise the velocities of CDM particles.

As mentioned in the introduction, a number of 
authors have previously noticed that tangential motions 
of particles can cause flattening in the inner density profile 
of dark matter halos (e.g. White \& Zartisky 1992; Ryden 1993; 
Sikivie \etal 1997; Subramanian \etal 2000; Subramanian 2000;
Hiotelis 2002; Le Delliou \& Henriksen 2003; Shapiro \etal 2004; 
Barnes \etal 2005).  Applying the collisionless Boltzmann equation to 
self-similar gravitational collapse, Subramanian (2000) and
Subramanian \etal (2000) demonstrated that tangential velocities 
are required to obtain an inner profile that is 
shallower than that of a singular isothermal sphere. They
suggested that an appropriate mixture of radial and tangential 
velocities may produce an inner $r^{-1}$ profile.
However, these authors did not propose a specific model 
that predicts such a profile. Furthermore, as we have shown 
in this paper, isotropic velocity dispersion alone 
is not sufficient for generating an inner $r^{-1}$ profile; 
rapid accretion with a quickly deepening potential well is also 
required in our model. Thus, although our explanantion about 
the inner density profile of dark matter halos is related to those 
presented in these earlier analyses, it provides additional physical 
insight into the problem, in particular in connection to
the properties of the mass accretion history that influence the 
final density profile.

Since our model only relies on the mass accretion history and orbit
isotropisation to explain the origin of NFW profiles, it naturally
explains why collapses with artificially reduced substructure
(Moore \etal 1999) also lead to the universal profile, even when the 
initial condition is as smooth as three intersecting plane
waves (Shapiro \etal 2004). However, if it is so smooth that the orbits
are not isotropised then a steeper central slope may result.
Note that although our explanation depends on different mass accretion
histories it does not explicitly depend on the dynamical friction and tidal
distribution of the substructures that makes up the actual 
mass accretion in cosmological $N$-body simulations (e.g. Dekel \etal 2003).

It is remarkable that our one-dimensional simulations with their two
simple ingredients reproduce the structural features of the full
three-dimensional simulations. In retrospect, this result is
consistent with the physical nature of hierarchical formation.
Because dark-matter halos are extended, even equal mass mergers are
relatively quiescent, their mutual orbits slowly decaying by
dynamical friction and ending with a low velocity merger, tidal
dissolution, and phase mixing.  Such events are likely to produce
sufficient scattering to yield isotropisation but not as violent as
envisioned by Lynden-Bell (1967). Therefore, once we have the two
necessary ingredients, the final equilibrium profile may not be
sensitive to the exact way in which the system settles into its final
equilibrium configuration.  Zhao et al. (2003a) show that a
significant correlation still exists between the final and initial
binding energies even for particles that are accreted in the fast
accretion phase based on 5 halos in their high-resolution simulations.
Reinforcing this point, we also find that the final energy of a
particle is correlated with its initial energy, and hence the
energy distribution of particles is determined by the initial 
conditions rather than by complete relaxation. The match between our
one-dimensional model and the three-dimensional cosmological
simulations suggests that violent relaxation might not play an
important role in redistributing the energies of particles except 
through isotropising the velocity field.

%%%%%%%%%%%%%%%
% Acknowledgments
%%%%%%%%%%%%%%%

\section*{Acknowledgments}
We thank Yun Li for help in generating mass accretion histories.  HJM
thanks Saleem Zaroubi for interesting discussions related to the
present work while both of them were at the Max-Planck-Institut f\"ur
Astrophysik, Garching, Germany.  NK and MDW would like to acknowledge
the support of NASA ATP NAG5-12038 \& NAGS-13308 and NSF AST-0205969.

%%%%%%%%%%%%%%%
% Bibliography
%%%%%%%%%%%%%%%

\label{lastpage}


\begin{thebibliography}{}
%\bibitem[]{Al02}
%Alvarez M.A., Ahn K., Shapiro P.R., 2003, RevMexAA SC, 18, 4
\bibitem[]{As04}
Ascasibar Y., Yepes G., Gottl\"ober S., M\"uller V., 2004, MNRAS, 352, 1109
\bibitem[]{Av98}
Avila-Reese V., Firmani C., Hern\'andez X., 1998, ApJ, 505, 37
\bibitem[]{Ba05}
Barnes E.I., Williams L.L.R., Babul A., Dalcanton J.J., 2005, ApJ, 634, 775
\bibitem[]{Be85}
Bertschinger E., 1985, ApJS, 58, 39
%\bibitem[]{Br98}
%Bryan G.L., Norman M., 1998, ApJ, 495, 80
\bibitem[]{Bu01}
Bullock J.S., Kolatt T.S., Sigad Y., Somerville R.S., Kravtsov A.V., Klypin A.A., Primack J.R., Dekel A., 2001, MNRAS, 321, 559
\bibitem[Carroll \etal(1992)]{Ca92}
Carroll S. M., Press W. H., Turner E. L. 1992, ARAA, 30, 499
\bibitem[]{Co00}
Col\'{i}n P., Klypin A.A., Kravtsov A.V., 2000, ApJ, 539, 561
\bibitem[]{De03}
Dekel A., Arad I, Devor J., Birnboim Y, 2003, ApJ, 588, 680
\bibitem[]{Di05}
Diemand J., Zemp M., Moore B., Stadel J., Carollo M., 2005, MNRAS, 364, 665
\bibitem[]{Ek98}
Eke V.R., Navarro J.F., Frenk C.S., 1998, ApJ, 503, 569
\bibitem[]{Ek01}
Eke V.R., Navarro J.F.,  Steinmetz M., 2001, ApJ, 554, 114
\bibitem[]{Fi84}
Fillmore J.A., Goldreich P., 1984, ApJ, 281, 1
\bibitem[]{Fu97}
Fukushige T., Makino J., 1997, ApJ, 477, L9
\bibitem[]{Fu01}
Fukushige T., Makino J., 2001, ApJ, 557, 533
\bibitem[]{Fu03}
Fukushige T., Makino J., 2003, ApJ, 588, 674
\bibitem[]{Gh00}
Ghigna S., Moore B., Governato F., Lake G., Quinn T., Stadel J., 2000, ApJ, 544, 616
\bibitem[]{Go75}
Gott J.R., 1975, ApJ, 201, 296
\bibitem[]{Gu77}
Gunn J.E., 1977, ApJ, 218, 592
\bibitem[]{Gu72}
Gunn J.E., Gott J.R., 1972, ApJ, 176, 1
\bibitem[]{Ha04}
Hansen S.H., Moore B., 2004, preprint, astro-ph/0411473
\bibitem[]{Hi02}
Hiotelis N., 2002, A\&A, 382, 84
\bibitem[]{Ho85}
Hoffman Y., Shaham J., 1985, ApJ, 297, 16
\bibitem[]{Hu99}
Huss A., Jain B., Steinmetz M., 1999, ApJ, 517, 64
\bibitem[]{Jing00a}
Jing Y.P., 2000, ApJ, 535, 30
\bibitem[]{Jing00b}
Jing Y.P., Suto Y., 2000, ApJ, 529, L69
\bibitem[]{Kl01}
Klypin A., Kravtsov A.V., Bullock J.S., Primack J.R., 2001, ApJ, 554, 903
\bibitem[]{LH03}
Le Delliou M., Henriksen R.N., 2003, A\&A, 408, 27
\bibitem[]{Li05}
Li Y., Mo H.J., vand den Bosch F.C., 2005, preprint, astro-ph/0510372
\bibitem[]{Ly67}
Lynden-Bell D., 1967, MNRAS, 136, 101
\bibitem[]{Mon02}
Monaco P., Theuns T., Taffoni G., 2002, MNRAS, 331, 587
\bibitem[]{Moore99}
Moore B., Quinn, T., Governato F., Stadel J., Lake G., 1999, MNRAS, 310, 1147
\bibitem[]{NFW96}
Navarro J.F., Frenk C.S., White S.D.M., 1996, ApJ, 462, 563
\bibitem[]{NFW97}
Navarro J.F., Frenk C.S., White S.D.M., 1997, ApJ, 490, 493
\bibitem[]{Nu01}
Nusser A., 2001, MNRAS, 325, 1397
\bibitem[]{Qu97}
Quinn T., Katz N., Stadel J., Lake G., 1997, preprint, astro-ph/9710043
\bibitem[]{Ri03}
Ricotti M., 2003, MNRAS, 344, 1237
\bibitem[]{Ry88}
Ryden B.S., 1988, ApJ, 333, 78
\bibitem[]{Ry93}
Ryden B.S., 1993, ApJ, 418, 4
\bibitem[]{Ry87}
Ryden B.S., Gunn J.E., 1987, ApJ, 318, 15
\bibitem[]{Shapiro04}
Shapiro P.R., Iliev I.T., Martel H., Ahn K., Avarez M.A., 2004, preprint, astro-ph/0409173
\bibitem[]{SI97}
Sikvie P., Tkachev I.I., Wang Y., 1997, Phys. Rev. D, 56, 1863
\bibitem[]{Spr05}
Springel V., 2005, MNRAS, 364, 1105
\bibitem[]{Su00}
Subramanian K., 2000, ApJ, 538, 517
\bibitem[]{SCO00}
Subramanian K., Cen R., Ostriker J.P., 2000, ApJ, 538, 528
\bibitem[]{Ta01}
Taylor J.E., Navarro J.F., 2001, ApJ, 563, 483
\bibitem[]{Tasitsiomi04}
Tasitsiomi A., Kravtsov A.V., Gottloeber S., Klypin A.A., 
  2004, ApJ, 607, 125
\bibitem[]{THL-B86}
 Tremaine S., Henon M., Lynden-Bell D., 1986, MNRAS, 219, 285 
\bibitem[]{va82}
van Albada T.S., 1982, MNRAS, 201, 939
\bibitem[]{Bo02}
van den Bosch F. C., 2002, MNRAS, 331, 98
\bibitem[]{We02}
Wechsler R.H., Bullock J.S., Primack J.R., 
     Kravtsov A.V., Dekel A., 2002, ApJ, 568, 52
\bibitem[]{WZ92}
White S. D. M., Zaritsky D., 1992, ApJ, 394, 1
\bibitem[]{Z93}
Zaroubi S., Hoffman Y., 1993, ApJ, 416, 410
\bibitem[]{Z03}
Zhao D., Mo H.J., Jing Y.P., B\"orner G., 2003a, MNRAS, 339, 12
\bibitem[]{Z04}
Zhao D., Jing Y.P., Mo H.J., B\"orner G., 2003b, ApJ, 597, L9


\end{thebibliography}
\end{document}